\documentclass[showpacs,preprintnumbers,amsmath,amssymb,twocolumn]{revtex4}

\usepackage{graphicx}
\usepackage{dcolumn}
\usepackage{bm}

\begin{document}

\title{Relativistic BCS-BEC Crossover at Finite Temperature\\ and Its Application to Color Superconductivity}
\author{\normalsize{Lianyi He and Pengfei Zhuang}}
\affiliation{Physics Department, Tsinghua University, Beijing
100084, China}

\begin{abstract}
The non-relativistic $G_0 G$ formalism of BCS-BEC crossover at
finite temperature is extended to relativistic fermion systems.
The uncondensed pairs contribute a pseudogap to the fermion
excitations. The theory recovers the BCS mean field approximation
at zero temperature and the non-relativistic results in a proper
limit. For massive fermions, when the coupling strength increases,
there exist two crossovers from the weak coupling BCS superfluid
to the non-relativistic BEC state and then to the relativistic BEC
state. For color superconductivity at moderate baryon density, the
matter is in the BCS-BEC crossover region, and the behavior of the
pseudogap is quite similar to that found in high temperature
superconductors.
\end{abstract}

\date{\today}

\maketitle

\section {Introduction}
\label{s1}
It is well-known that, by adjusting the attractive coupling
strength among the constituents, a fermion system may undergo a
smooth crossover from the Bardeen-Cooper-Shriffer (BCS)
superfluidity/superconductivity in degenerate fermion gas to the
Bose-Einstein condensation (BEC) of composite molecules. Such a
BCS-BEC crossover is theoretically due to the fact that the wave
functions of BCS and BEC ground states are essentially the
same\cite{BCSBEC1,BCSBEC2}. The BCS-BEC crossover is expected to
be realized in high temperature superconductor and atomic fermion
gas\cite{BCSBEC3,BCSBEC4,BCSBEC5,BCSBEC6,BCSBEC7,BCSBEC8} via
using an external magnetic field to change the s-wave scattering
length\cite{exp}.

The superconductivity in Quantum Chromodynamics (QCD), i.e., the
color superconductivity\cite{CSCreview}, is naturally considered
as a system to study the relativistic BCS-BEC crossover. Due to
the asymptotic property of QCD, there may exist a crossover from
the BCS superconductivity with weakly bound quark pairs at high
baryon density to the BEC state of composite hadrons at low baryon
density\cite{qq}. Such a BCS-BEC crossover in QCD may also be
realized in chiral condensed matter\cite{qqbar,qqbar2} and in pion
superfluid\cite{qq}. At moderate baryon density, while a diquark
BEC state may not be realized due to the chiral symmetry
restoration, the attractive coupling strength is obviously not
located in the weak coupling region. It is shown in many effective
QCD models that, the quark energy gap at moderate baryon density
is about $100$ MeV\cite{CSCgap} which is already of the order of
the Fermi energy. The strong coupling in this case may induce a
so-called pseudogap effect, which has been investigated in two
flavor color superconductivity above the critical
temperature\cite{CSCpg}. A natural question is how the pseudogap
modifies the critical temperature and thermodynamics of the color
superconductor. To answer this question, one needs to construct a
relativistic theory at finite temperature which can describe the
pseudogap and possible BCS-BEC crossover.

The BCS-BEC crossover in relativistic fermion systems are recently
investigated in the Nozieres--Schmitt-Rink (NSR) theory above the
critical temperature\cite{RBCSBEC1,RBCSBEC2}, the boson-fermion
model\cite{RBCSBEC3} and the BCS-Leggett mean field theory at zero
temperature\cite{RBCSBEC4}. It is shown that, not only the BCS
superfluidity and the non-relativistic BEC (NBEC) of heavy
molecules but also the NBEC and the relativistic BEC (RBEC) of
nearly massless molecules can be smoothly connected. In the RBEC
state, anti-fermion pairs (anti-bosons) are excited and become
nearly degenerate with fermion pairs (bosons). From the NSR theory
at $T\ge T_c$, where $T_c$ is the critical temperature, the
difference between the NBEC\cite{kerson} and
RBEC\cite{kapusta,RBEC} states is
significant\cite{RBCSBEC1,RBCSBEC2}.

It is widely known that, at zero temperature the mean field theory
is a good approximation to describe the BCS-BEC
crossover\cite{BCSBEC9}, and the pair fluctuations can be safely
neglected even at strong coupling. Only around the unitary limit,
i.e., the infinite scattering length limit, the pair fluctuations
are somewhat important to obtain a proper value of the universal
constant\cite{BCSBEC7}. In our previous paper\cite{RBCSBEC4}, we
investigated the generalization from non-relativistic to
relativistic BCS-BEC crossover at zero temperature in the
BCS-Leggett mean field theory. At finite temperature, however, the
condensed pairs with zero momentum can be thermally excited, and
one should go beyond the mean field approximation to treat
properly the uncondensed pairs\cite{BCSBEC6}.

There exist many methods to treat pair fluctuations at finite
temperature. In the NSR theory, which is also called $G_0G_0$
theory, the pair fluctuations enter only the number equation, and
the fermion loops which appear in the pair propagator are
constructed by bare Green function $G_0$. As a consequence, such a
theory is in principle not self-consistent and is valid only at
$T\geq T_c$. For the study of BCS-BEC crossover, one needs a
theory which is valid not only above the critical temperature but
also in the symmetry breaking phase. While such a strict theory is
not yet reached so far, some T-matrix approaches are recently
developed, see for instance \cite{BCSBEC6,BCSBEC9}. Among them,
the asymmetric pair approximation or the so-called $G_0G$
scheme\cite{BCSBEC6,G0G} is a competitive one. The effect of the
pair fluctuations in the $G_0G$ method is treated as a fermion
pesudogap which has been widely discussed in high temperature
superconductivity.  In contrast to the NSR theory ($G_0 G_0$
scheme), the $G_0 G$ scheme is self-consistent and keeps the Ward
identity\cite{BCSBEC6}.

In the study of color superconductivity at moderate density, the
color condensed phase is of great interest. The NSR
theory\cite{RBCSBEC1,RBCSBEC2}, which seems valid in the normal
phase, can only predict the transition temperature of color
superconductivity. A necessary task in this field of research is
to develop a relativistic BCS-BEC crossover theory in the symmetry
breaking phase. In this paper, we will generalize the $G_0 G$
scheme to relativistic fermion systems. A necessary requirement
for such a generalization is to recover the non-relativistic
limit\cite{BCSBEC6} and mean field limit\cite{BCSBEC4} properly.
With this theory, we can calculate the critical temperature $T_c$
for arbitrary coupling and describe the BCS-NBEC-RBEC crossover at
finite temperature. It, as an application, can be used to study
the pseudogap effect on color superconductivity.

The paper is organized as follows. In section \ref{s2} we review
the BCS mean field theory for relativistic
superfluidity/superconductivity. In the framework of the $G_0 G$
scheme, we include in Section \ref{s3} the contribution from the
uncondensed pairs and construct coupled equations for the
superfluid order parameter and pseudogap. In section \ref{s4}, we
apply the theory to massive fermions and study the BCS-NBEC-RBEC
crossover at finite temperature. In section \ref{s5}, we apply the
theory to color superconducting quark matter. We will calculate
the transition temperature and the quark pseudogap and show the
significance of the fluctuations at moderate baryon density. We
summarize in section \ref{s6}.

\section {BCS Mean Field Theory}
\label{s2}
We consider a model with only fermions as elementary blocks. The
Lagrangian density can be written as
\begin{equation}
{\cal L}=\bar{\psi}\left(i\gamma^\mu\partial_\mu-m\right)\psi
+{\cal L}_I ,
\end{equation}
where $\psi,\bar{\psi}$ denote the Dirac fermion fields with mass
$m$, and ${\cal L}_I$ indicates the attractive interaction among
fermions. Since the dominant interaction is the $J^P=0^+$ scalar
channel, the interaction for the pairing between different spins
can take the form\cite{RBCSBEC1,RBCSBEC4}
\begin{equation}
{\cal L}_I=\frac{g}{4}\left(\bar{\psi} i\gamma_5C\bar{\psi}^{\text
T}\right)\left(\psi^{\text T}C i\gamma_5\psi\right) ,
\end{equation}
where $g$ is the attractive coupling constant, and
$C=i\gamma_0\gamma_2$ is the charge conjugation matrix. Generally,
by adjusting the coupling strength, the crossover from
condensation of spin-zero Cooper pairs with large size at weak
coupling to the Bose-Einstein condensation of deeply bound bosons
at strong coupling can be realized. In our model, only fermions
are elementary particles. Another type of model which is used to
discuss the BCS-BEC crossover in high temperature superconductors
and atomic Fermi gases is the so-called boson-fermion model where
both fermions and bosons are considered as elementary blocks. Such
a model is recently generalized to study the relativistic BCS-BEC
crossover\cite{RBCSBEC3}.

In order to develop a finite temperature theory including pair
fluctuations in the symmetry breaking phase, we first review in
this section the BCS mean field theory in the functional integral
approach and $G_0 G$ formalism.

\subsection {Functional Integral Approach}
In the functional integral approach, we start the calculation from
the partition function in imaginary time formalism,
\begin{equation}
Z=\int D\bar{\psi} D\psi e^{\int_0^\beta d\tau\int d^3{\bf
x}({\cal L}+\mu\psi^\dagger\psi)}
\end{equation}
where $\beta$ is the inverse temperature, $\beta=1/T$, and $\mu$
is the chemical potential corresponding to the net charge density
$\psi^\dagger\psi$ and is determined by the charge conservation.
Performing a Hubbard-Stratonovich transformation which introduces
an auxiliary pair field $\Delta(x)=g\psi^T(x)Ci\gamma_5\psi(x)/2$,
and then integrating out the fermions, we derive the partition
function
\begin{equation}
Z=\int D\Delta D\Delta^*\ e^{-S_{\text{eff}}[\Delta,\Delta^*]}
\end{equation}
with the effective boson action
\begin{equation}
\label{eff}
S_{\text{eff}}=\int_0^\beta d\tau\int d^3{\bf
x}\left[\frac{|\Delta(x)|^2}{g}-\frac{1}{2\beta}\text{Tr}\ln
[\beta{\bf G}^{-1}]\right]
\end{equation}
in terms of the inverse Nambu-Gorkov propagator
\begin{eqnarray}
\label{g} {\bf
G}^{-1}=i\gamma^\mu\partial_\mu-m+\mu\gamma_0\sigma_3+i\gamma_5\Delta\sigma_++i\gamma_5\Delta^*\sigma_-,
\end{eqnarray}
where $\sigma_\pm=(\sigma_1\pm i\sigma_2)/2$ are defined in the
Nambu-Gorkov space with $\sigma_i(i=1,2,3)$ being the Pauli
matrices.

The mean field theory is a good approximation to describe the
BCS-BEC crossover at low enough temperature, namely $T\ll T_c$,
since the dominant contribution of fluctuations to the effective
potential is from the Goldstone mode and is proportional to
$T^4$\cite{BCSBEC5}. In the mean field approximation, we consider
a uniform static saddle point $\Delta(x)=\Delta_{\text{sc}}$ which
satisfies the stationary condition $\delta
S_\text{eff}[\Delta_{\text{sc}}]/\delta \Delta_{\text{sc}}=0$. The
thermodynamic potential
$\Omega_{\text{mf}}=S_\text{eff}[\Delta_{\text{sc}}]/(\beta V)$ at
the saddle point  can be evaluated as
\begin{eqnarray}
\Omega_\text{mf} &=&{\Delta_\text{sc}^2\over g}-\int{d^3 {\bf
k}\over (2\pi)^3}\Big[\left(E_{\bf k}^++E_{\bf k}^--\xi_{\bf
k}^+-\xi_{\bf k}^-\right)\nonumber\\
&&-{1\over \beta}\left(\ln(1+e^{-\beta E_{\bf
k}^+})+\ln(1+e^{-\beta E_{\bf k}^-})\right)\Big],
\end{eqnarray}
where we have defined the quasi-particle energies $E_{\bf
k}^\pm=\sqrt{(\xi_{\bf k}^\pm)^2+\Delta_{\text{sc}}^2}$ with $
\xi_{\bf k}^\pm=\epsilon_{\bf k}\pm\mu$ and $\epsilon_{\bf
k}=\sqrt{{\bf k}^2+m^2}$. Minimizing $\Omega_{\text{mf}}$, we get
the gap equation to determine the order parameter
$\Delta_\text{sc}$ in the symmetry breaking phase,
\begin{equation}
\label{gap1} \frac{1}{g}=\int{d^3{\bf k}\over
(2\pi)^3}\left[\frac{1-2f(E_{\bf k}^-)}{2E_{\bf
k}^-}+\frac{1-2f(E_{\bf k}^+)}{2E_{\bf k}^+}\right],
\end{equation}
where $f(x)=1/(e^{\beta x}+1)$ is the Fermi-Dirac distribution
function. In the study of BCS-BEC crossover, people often
consider the thermodynamics in canonical ensemble with fixed
fermion density $n$ by fixing the Fermi momentum $k_f$ through the
relation $n=k_f^3/(3\pi^2)$ at zero temperature. At finite
temperature, the density can be obtained from the first order
derivative of the thermodynamic potential with respect to the
chemical potential,
\begin{eqnarray}
\label{number}
n&=&\int{d^3{\bf k}\over
(2\pi)^3}\bigg[\left(1-\frac{\xi_{\bf k}^-}{E_{\bf
k}^-}(1-2f(E_{\bf
k}^-))\right)\nonumber\\
&&-\left(1-\frac{\xi_{\bf k}^+}{E_{\bf k}^+}(1-2f(E_{\bf
k}^+))\right)\bigg].
\end{eqnarray}
The first and second terms in the square bracket on the right hand
side of equations (\ref{gap1}) and (\ref{number}) correspond
respectively to fermion and anti-fermion degrees of freedom.

\subsection {$G_0 G$ Formalism}
Now we reexpress the BCS mean field theory in the $G_0 G$
formalism\cite{BCSBEC6,BCSBEC7,TBCS}. Such a formalism is
convenient for us to go beyond the BCS and include uncondensed
pairs at finite temperature. Let us start from the fermion
propagator ${\cal S}$ in the symmetry breaking phase. The inverse
propagator reads
\begin{equation}
{\cal S}^{-1}(k)= \left(\begin{array}{cc} {\cal G}_0^{-1}(k,\mu)&i\gamma_5\Delta_{\text{sc}}\\
i\gamma_5\Delta_{\text{sc}}&{\cal
G}_0^{-1}(k,-\mu)\end{array}\right)
\end{equation}
with the inverse free propagator
\begin{equation}
\label{g0}
{\cal G}_0^{-1}(k,\mu) = (i\omega_n+\mu)\gamma_0-{\bf
\gamma}\cdot{\bf k}-m,
\end{equation}
where $k=(i\omega_n,{\bf k})$ is the fermion four momentum at
finite temperature with $\omega_n$ being the fermion frequency
$\omega_n=(2n+1)\pi T$ $(n=0,\pm 1, \pm 2,\cdots)$. The propagator
can be formally expressed as
\begin{equation}
{\cal S}(k)= \left(\begin{array}{cc} {\cal G}(k,\mu)&{\cal F}(k,\mu)\\
{\cal F}(k,-\mu)&{\cal G}(k,-\mu)\end{array}\right)
\end{equation}
with the diagonal and off-diagonal elements
\begin{eqnarray}
{\cal G}(k,\mu)&=&\left[{\cal
G}_0^{-1}(k,\mu)-\Sigma_{\text{sc}}(k)\right]^{-1},\nonumber\\
{\cal F}(k,\mu)&=&-{\cal G}(k,\mu)i\gamma_5\Delta_{\text{sc}}{\cal
G}_0(k,-\mu),
\end{eqnarray}
where the fermion self-energy $\Sigma_\text{sc}$ is defined as
\begin{eqnarray}
\Sigma_\text{sc}(k)&=&i\gamma_5\Delta_\text{sc}{\cal G}_0
(k,-\mu)i\gamma_5\Delta_{\text{sc}}\nonumber\\
&=&-\Delta_\text{sc}^2{\cal G}_0(-k,\mu).
\end{eqnarray}
With the help of the energy projectors
\begin{equation}
\Lambda_{\pm}({\bf k}) = {1\over
2}\left[1\pm{\gamma_0\left(\vec{\gamma}\cdot{\bf k}+m\right)\over
\epsilon_{\bf k}}\right],
\end{equation}
the propagator elements can be explicitly evaluated as
\begin{eqnarray}
\label{element}
{\cal G}(k,\mu)&=& {\left(i\omega_n+\xi_{\bf
k}^-\right)\Lambda_+\gamma_0\over
(i\omega_n)^2-(E_{\bf k}^-)^2}+ {\left(i\omega_n-\xi_{\bf k}^+
\right)\Lambda_-\gamma_0\over (i\omega_n)^2-(E_{\bf k}^+)^2},\nonumber\\
{\cal F}(k,\mu)&=& {i\Delta_\text{sc}\Lambda_+\gamma_5\over
(i\omega_n)^2-(E_{\bf k}^-)^2}+
{i\Delta_\text{sc}\Lambda_-\gamma_5\over (i\omega_n)^2-(E_{\bf
k}^+)^2}.
\end{eqnarray}
The gap equation for the order parameter $\Delta_\text{sc}$ is
related to the off-diagonal element,
\begin{eqnarray}
\label{gap2}
\Delta_\text{sc}&=&-i\frac{g}{2}\sum_k\text{Tr}\left[i\gamma_5{\cal
F}(k,\mu)\right]\nonumber\\
&=&-i\frac{g}{2}\Delta_{\text{sc}}\sum_k\text{Tr}\left[{\cal
G}(k,\mu){\cal G}_0(-k,\mu)\right],
\end{eqnarray}
and the fermion number is controlled by the diagonal element,
\begin{equation}
n=-i\sum_k\text{Tr}\left[\gamma_0{\cal G}(k,\mu)\right]
\end{equation}
with the four momentum integration $\sum_k=iT\sum_n\int d^3{\bf
k}/\left(2\pi\right)^3$ at finite temperature. Completing the
Matsubara frequency summation, we can reobtain the gap equation
(\ref{gap1}) and number equation (\ref{number}).

In the BCS mean field theory, fermion--fermion pairs and
anti-fermion--anti-fermion pairs explicitly enter the system below
$T_c$ only through the condensate $\Delta_\text{sc}$. In the $G_0
G$ formalism, the fermion self-energy can equivalently be
expressed as
\begin{equation}
\Sigma_\text{sc}(k)=\sum_q t_\text{sc}(q){\cal G}_0(q-k,\mu)
\end{equation}
associated with a condensed-pair propagator given by
\begin{equation}
t_\text{sc}(q)=i\frac{\Delta_\text{sc}^2}{T}\delta(q),
\end{equation}
where $q=(i\nu_n,{\bf q})$ is the boson four momentum with boson
frequency $\nu_n=2n\pi T$.

The BCS theory can be related to a specific pair susceptibility
$\chi$ defined by
\begin{equation}
\chi_\text{BCS}(q)=-\frac{i}{2}\sum_k\text{Tr}\left[{\cal
G}(k,\mu){\cal G}_0(q-k,\mu)\right],
\end{equation}
with which, the gap equation for the condensate $\Delta_\text{sc}$
can be written as
\begin{equation}
1-g\chi_{\text{BCS}}(0)=0.
\end{equation}
This implies that the uncondensed pair propagator should be of the
form
\begin{equation}
t(q)=\frac{ig}{1-g\chi_{\text{BCS}}(q)},
\end{equation}
and $t^{-1}(q=0)$ is proportional to the pair chemical potential
$\mu_{\text{pair}}$. Therefore, the fact that in the symmetry
breaking phase the pair chemical potential is zero leads to the
BEC-like condition
\begin{equation}
t^{-1}(q=0)=0.
\end{equation}

While the uncondensed pairs do not play any real role in the BCS
mean field theory, such a specific choice of the pair
susceptibility and the BEC-like condition tell us a way how to go
beyond the BCS mean field theory and include the effect of
uncondensed pairs.

\section {Beyond Mean Field Theory}
\label{s3}
\begin{figure}
\centerline{\includegraphics[]{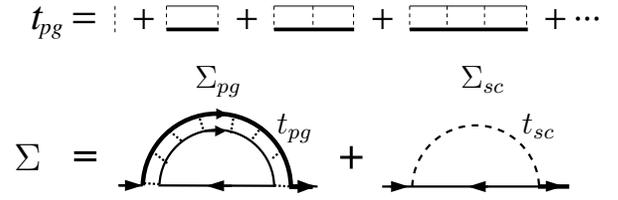}} \caption{Diagramatic
representation of the propagator $t_\text{pg}$ for the uncondensed
pairs and the fermion self-energy. The total fermion self-energy
contains contributions from condensed ($\Sigma_\text{sc}$) and
uncondensed ($\Sigma_\text{pg}$) pairs. The dashed, thin solid and
thick solid lines in $t_\text{pg}$ represent, respectively, the
coupling constant $g/2$, bare propagator ${\cal G}_0$ and full
propagator ${\cal G}$.  This diagram is taken from \cite{diagram}.
\label{fig1}}
\end{figure}

While the uncondensed pairs can be safely neglected at weak
coupling, they should be included for a self-consistent theory at
arbitrary coupling and at finite temperature. We now go beyond the
BCS mean field approximation and include the uncondensed pairs in
the $G_0 G$ formalism. It is clear that, in the BCS mean field
approximation the fermion self-energy $\Sigma_\text{sc}$ includes
contribution only from the condensed pairs. At finite temperature,
the condensed pairs with zero total momentum can be thermally
excited, and the total propagator should contain both the
condensed (sc) and uncondensed or ``pseudogap"-associated (pg)
contributions,
\begin{eqnarray}
t(q)&=&t_\text{pg}(q)+t_\text{sc}(q),\nonumber\\
t_\text{pg}(q)&=&\frac{ig}{1-g\chi(q)},\ \ \ q\neq 0,\nonumber\\
t_\text{sc}(q)&=&i\frac{\Delta_\text{sc}^2}{T}\delta(q).
\end{eqnarray}
Now the total fermion self-energy becomes
\begin{equation}
\Sigma(k)=\sum_q t(q){\cal
G}_0(q-k,\mu)=\Sigma_\text{sc}(k)+\Sigma_\text{pg}(k),
\end{equation}
with the mean field part
\begin{equation}
\Sigma_\text{sc}(k)=\sum_q t_\text{sc}(q){\cal G}_0(q-k,\mu)
\end{equation}
and the pseudogap related part
\begin{equation}
\Sigma_\text{pg}(k)=\sum_q t_\text{pg}(q){\cal G}_0(q-k,\mu).
\end{equation}
With the full propagator
\begin{equation}
{\cal G}(k,\mu)=\left[{\cal G}_0^{-1}(k,\mu)-\Sigma(k)\right]^{-1}
\end{equation}
in terms of the total self-energy, the pair susceptibility is
still given by
\begin{equation}
\label{sus} \chi(q)=-\frac{i}{2}\sum_k \text{Tr}{\cal
G}(k,\mu){\cal G}_0(q-k,\mu).
\end{equation}
The $G_0 G$ formalism used here is diagrammatically illustrated in
Fig.\ref{fig1}.

Note that, the feedback of the pair fluctuations on the order
parameter $\Delta_\text{sc}$ is included, and it and the chemical
potential $\mu$ are in principle determined by the BEC condition
$t_\text{pg}^{-1}(0)=0$ and the number equation
$n=-i\sum_k\text{Tr}\left[\gamma_0{\cal G}(k,\mu)\right]$.

The above equations are hard to handle analytically. In the
symmetry breaking phase with $T\leq T_c$, the BEC condition
$t_\text{pg}^{-1}(0)=0$ implies that $t_\text{pg}(q)$ is peaked at
$q=0$. This allows us to approximate
\begin{equation}
\label{appro}
\Sigma(k)\simeq -\Delta^2 {\cal G}_0(-k,\mu),
\end{equation}
where $\Delta^2$ contains contributions from the condensed and
uncondensed pairs,
\begin{equation}
\Delta^2=\Delta_\text{sc}^2+\Delta_\text{pg}^2
\end{equation}
with the pseudogap $\Delta_\text{pg}$ defined as
\begin{equation}
\label{pg1}
\Delta_\text{pg}^2 = -\sum_{q\neq 0}t_\text{pg}(q).
\end{equation}
It is necessary to point out that, above the critical temperature
$T_c$ such an approximation is no longer good, since the BEC
condition is not valid in normal phase.

Since the pseudogap $\Delta_\text{pg}$ looks similar to the
condensate $\Delta_\text{sc}$, a natural question is whether a
finite $\Delta_\text{pg}$ breaks the symmetry of the system. If
yes, $\Delta_\text{sc}$ will no longer be considered as the order
parameter of the phase transition. By omitting a term of the order
of $O\left(\Delta_\text{sc}^2/\Lambda^2\right)$, where $\Lambda$
is a momentum cutoff, the inverse fermion propagator including the
feedback of the pair fluctuations can be written as
\begin{equation}
{\cal S}^{-1}(k)= \left(\begin{array}{cc} {\cal G}_0^{-1}(k,\mu)-\Sigma_\text{pg}(k)&i\gamma_5\Delta_{\text{sc}}\\
i\gamma_5\Delta_{\text{sc}}&{\cal
G}_0^{-1}(k,-\mu)-\Sigma_\text{pg}^\prime(k)\end{array}\right)
\end{equation}
where
$\Sigma_\text{pg}^\prime=\Sigma_\text{pg}(\mu\rightarrow-\mu)$. It
is now clear that, the pseudogap appears in the diagonal elements
of the Nambu-Gorkov propagator and does not break the symmetry of
the system. On the other hand, parallel to the discussion in
non-relativistic theory\cite{BCSBEC6,BCSBEC7,G0G}, we can show
that $\Delta_\text{pg}^2$ is just the fluctuation of the order
parameter field $\Delta(x)$,
\begin{equation}
\Delta_\text{pg}^2=\langle
|\Delta|^2\rangle-\langle|\Delta|\rangle^2,
\end{equation}
and hence it does not break the symmetry.

Under the approximation (\ref{appro}), all the equations in the
mean field theory are still valid, the only change is the
replacement of $E_{\bf k}^\pm=\sqrt{\left(\xi_{\bf
k}^\pm\right)^2+\Delta_\text{sc}^2}$ by $E_{\bf
k}^\pm=\sqrt{\left(\xi_{\bf k}^\pm\right)^2+\Delta^2}$. For
instance, the diagonal element ${\cal G}$ of the full propagator,
the fermion number $n$ and the gap equation for $\Delta$ take
exactly their mean field forms (\ref{element}), (\ref{number}) and
(\ref{gap1}). The equations (\ref{gap1}), (\ref{number}) and
(\ref{pg1}) determine self-consistently the order parameter
$\Delta_\text{sc}$, the pseudogap $\Delta_\text{pg}$ and the
chemical potential $\mu$ as functions of temperature $T$. Note
that the pair fluctuation effect is self-consistently included in
the coupled equations through the pseudogap $\Delta_{\text{pg}}$.
It is necessary to point out that, the $G_0 G$ approach we used is
quite different from the NSR theory. In the NSR theory, the pair
fluctuations enter only the number equation via adding a molecule
number term\cite{RBCSBEC1,RBCSBEC2}.

However, solving such a coupled set of equations is still rather
complicated. Fortunately, the BEC condition allows us to do
further approximations for the pair propagator $t_{\text{pg}}(q)$.
Using the BEC condition $1-g\chi(0)=0$, the T-matrix can be
written as
\begin{equation}
t_\text{pg}(q)=\frac{-i}{\chi(q)-\chi(0)}.
\end{equation}
Since the pseudogap is dominated by the gapless pair dispersion in
long wavelength limit, we can expand the susceptibility around
$q=0$ in this limit,
\begin{equation}
\label{tpg}
t_\text{pg}(q)\simeq\frac{-i}{Z_1q_0+Z_2q_0^2-\xi^2{\bf
q}^2+i\Gamma(q)},
\end{equation}
where the coefficients $Z_1,Z_2$ and $\xi^2$ are defined as
\begin{eqnarray}
Z_1&=&\frac{\partial\chi}{\partial q_0}\Big|_{q=0},\ \ \ \
Z_2=\frac{1}{2}\frac{\partial^2\chi}{\partial
q_0^2}\Big|_{q=0},\nonumber\\
\xi^2&=&-{1\over 2}\frac{\partial^2\chi}{\partial {\bf
q}^2}\Big|_{q=0},
\end{eqnarray}
and we have considered the fact that the susceptibility depends
only on ${\bf q}^2$. The explicit expressions for $Z_1, Z_2$ and
$\xi^2$ are listed in Appendix \ref{app}.

In the symmetry breaking phase where the temperature is low, it is
believed that the pairs are long-lived and we can neglect their
width $\Gamma$. With the expansion for the pair propagator, the
equation (\ref{pg1}) now takes a simple form
\begin{equation}
\label{pg2}
\Delta_{\text{pg}}^2=\frac{1}{Z_2}\int\frac{d^3{\bf
q}}{(2\pi)^3}\frac{1+b(\omega_{\bf q}-\nu)+b(\omega_{\bf
q}+\nu)}{2\omega_{\bf q}},
\end{equation}
where $b(x)=1/(e^{\beta x}-1)$ is the Bose-Einstein distribution
function and $\omega_{\bf q}$ and $\nu$ are defined as
\begin{equation}
\omega_{\bf q}=\sqrt{\nu^2+c^2{\bf q}^2},\ \ \
\nu=\frac{Z_1}{2Z_2},\ \ c^2=\frac{\xi^2}{Z_2}.
\end{equation}
The first term on the righthand side of (\ref{pg2}) suffers
ultraviolet divergence, but it can be dropped out via
renormalization\cite{BCSBEC6,BCSBEC7}.

Let us first discuss some conclusions from the above equations
without detailed numerical calculations.
\\ 1) At zero temperature, the pseudogap $\Delta_{\text{pg}}$
vanishes automatically and the theory is reduced to the BCS mean
field approximation\cite{RBCSBEC4}.
\\ 2) If the coupling is not so strong that the molecule binding energy
$E_b$ satisfies $E_b\ll 2m$, the theory is reduced to its
non-relativistic version\cite{BCSBEC6} for systems with $k_f\ll m$
or $n\ll m^3$.
\\ 3)If $Z_1$ dominates the propagator $t_\text{pg}$, the pair dispersion is quadratic in $|{\bf
q}|$, and the pseudogap $\Delta_\text{pg}$ can be analytically
integrated out and is proportional to $T^{3/4}$ at low
temperature. On the other hand, if $Z_2$ is the dominant term, the
pair dispersion is linear in $|{\bf q}|$ and $\Delta_{\text{pg}}$
becomes proportional to $T$ at low temperature. In the next
section, we will show that the first case happens in the NBEC
region and the second case occurs in the RBEC region.
\\ 4)From the explicit expression of $Z_1$ shown in Appendix \ref{app},
\begin{equation}
Z_1=\frac{1}{\Delta^2}\left[\frac{n}{2}-\int\frac{d^3{\bf
k}}{(2\pi)^3}\left(f(\xi_{\bf k}^-)-f(\xi_{\bf
k}^+)\right)\right],
\end{equation}
the quantity in the square brackets is just the total number
density $n_{\text{B}}$ of the bound pairs (bosons),
\begin{equation}
n_{\text{B}}=Z_1\Delta^2.
\end{equation}
From the relation
$\Delta^2=\Delta_{\text{sc}}^2+\Delta_{\text{pg}}^2$, $n_B$ can be
decomposed into the condensed pair number $n_{\text{sc}}$ and the
uncondensed pair number $n_{\text{pg}}$,
\begin{equation}
n_{\text{sc}}=Z_1\Delta_{\text{sc}}^2,\ \ \ \
n_{\text{pg}}=Z_1\Delta_{\text{pg}}^2.
\end{equation}
The fraction of the condensed pairs can be defined by
\begin{equation}
P_c=\frac{n_{\text{sc}}}{n/2}=\frac{2Z_1\Delta_{\text{sc}}^2}{n}.
\end{equation}
5) In the weak coupling BCS region, we expect the fermion number
density
\begin{equation}
n\simeq2\int\frac{d^3{\bf k}}{(2\pi)^3}\left(f(\xi_{\bf
k}^-)-f(\xi_{\bf k}^+)\right)
\end{equation}
which leads to $n_{\text{B}}=0$ in this region. In the deep BEC
region, however, almost all the fermions form two body bound
states which results in $n_{\text{B}}\simeq n/2$. At zero
temperature, we have $\Delta_\text{pg}=0,
n_{\text{B}}=n_{\text{sc}}$ and $P_c=1$, while at the critical
temperature $T_c$, the order parameter $\Delta_{\text{sc}}$
disappears, and the uncondensed pair number $n_{\text{pg}}$
becomes dominant and is approximately equal to $n/2$.

Numerically, the transition temperature $T_c$ can be calculated
from (\ref{pg2}) and the generalized equations (\ref{gap1}) and
(\ref{number}) by setting $\Delta_{\text{sc}}=0$. Usually, at and
above $T_c$ where the order parameter $\Delta_{\text{sc}}$
disappears, the pseudogap $\Delta_{\text{pg}}$ does not vanish. We
can define a limit temperature $T^*$ where the pseudogap starts to
disappear. Between the two temperatures $T_c$ and $T^*$ is the
so-called pseudogap phase. While the present $G_0 G$ formalism is
likely valid only in the symmetry breaking phase with $T\leq T_c$,
it can be generalized to the region above $T_c$ by introducing a
non-vanishing pair chemical potential
$\mu_{\text{pair}}$\cite{BCSBEC7}. We will do such a
generalization, but the numerical results in the following will be
presented mainly at $T\leq T_c$.

Above the critical temperature $T_c$, the order parameter
$\Delta_{\text{sc}}$ vanishes, and the BEC condition is no longer
valid, $1-g\chi(0)\neq0$. As a consequence, the propagator of the
pair takes the form
\begin{equation}
t_{\text{pg}}(q)=\frac{-i}{\chi(q)-\chi(0)-Z_0}
\end{equation}
with $Z_0=1/g-\chi(0)$. As an estimation of $\Delta_{\text{pg}}$
and $T^*$, we still perform the expansion for the susceptibility
around $q=0$,
\begin{equation}
t_{\text{pg}}(q)\simeq\frac{-i}{Z_1q_0+Z_2q_0^2-\xi^2|{\bf
q}|^2-Z_0+i\Gamma(q)}.
\end{equation}
Now the pseudogap equation becomes
\begin{equation}
\label{ppg}
\Delta_{\text{pg}}^2=\frac{1}{Z_2}\int\frac{d^3{\bf
q}}{(2\pi)^3}\frac{b(\omega^\prime_{\bf
q}-\nu)+b(\omega^\prime_{\bf q}+\nu)}{2\omega^\prime_{\bf q}}
\end{equation}
with the definition
\begin{equation}
\omega^\prime_{\bf q}=\sqrt{\nu^2+\lambda^2+c^2{\bf q}^2},\ \ \
\lambda^2=Z_0/Z_2.
\end{equation}
The equation (\ref{ppg}) together with the number equation
(\ref{number}) determines the pseudogap $\Delta_{\text{pg}}$ and
chemical potential $\mu$ above $T_c$. Since the pair dispersion is
now no longer gapless in the long-wavelength limit, and $Z_0$ will
generally increase with temperature, we expect that
$\Delta_{\text{pg}}$ will drop down and approach zero at the
dissociation temperature $T^*$.

In the end of this section, we discuss the thermodynamics of the
system. The naive BCS mean field theory does not include the
contribution from the uncondensed bosons which, however, dominate
the thermodynamics at strong coupling. In the present theory,
considering the uncondensed pairs, the total thermodynamic
potential $\Omega$ contains both the fermion and boson
contributions,
\begin{equation}
\Omega=\Omega_{\text{cond}}+\Omega_{\text{fermion}}+\Omega_{\text{boson}},
\end{equation}
where $\Omega_{\text{cond}}$ is from the condensed pairs,
\begin{equation}
\Omega_{\text{cond}} ={\Delta_{\text{sc}}^2\over g},
\end{equation}
$\Omega_{\text{fermion}}$ from the fermion excitations,
\begin{eqnarray}
\Omega_{\text{fermion}} &=&\int{d^3 {\bf k}\over
(2\pi)^3}\bigg[\left(\xi_{\bf k}^++\xi_{\bf k}^--E_{\bf
k}^+-E_{\bf k}^-\right)\\
&&-\frac{1}{\beta}\left(\ln{(1+e^{-\beta E_{\bf
k}^+})}+\ln{(1+e^{-\beta E_{\bf k}^-})}\right)\bigg],\nonumber
\end{eqnarray}
and $\Omega_{\text{boson}}$ from the uncondensed pairs,
\begin{equation}
\Omega_{\text{boson}}=\sum_q\ln[1-g\chi(q)].
\end{equation}
Under the approximation (\ref{tpg}) for the pair propagator, the
boson part in the symmetry breaking phase can be evaluated as
\begin{equation}
\Omega_{\text{boson}}=\frac{1}{\beta}\int{d^3 {\bf q}\over
(2\pi)^3}\left[\ln{(1-e^{-\beta \omega_{\bf
q}^+})}+\ln{(1-e^{-\beta \omega_{\bf q}^-})}\right]
\end{equation}
with $\omega_{\bf q}^\pm=\omega_{\bf q}\pm\nu$.

There exist two limiting cases for the boson contribution. If
$Z_1$ dominates the pair propagator, the pair dispersion is
quadratic in ${\bf q}$, and $\Omega_{\text{boson}}$ recovers the
thermodynamic potential of a non-relativistic boson gas,
\begin{equation}
\Omega_{\text{boson}}^{\text{NR}}=\frac{1}{\beta}\int{d^3 {\bf
q}\over (2\pi)^3}\ln\left(1-e^{-\beta {\bf q}^2/(2m_{\text
B})}\right).
\end{equation}
On the other hand, if $Z_2$ dominates the pair properties, the
pair dispersion is linear in $|{\bf q}|$ and we obtain the
thermodynamic potential for an ultra-relativistic boson gas
\begin{equation}
\Omega_{\text{boson}}^{\text{UR}}=\frac{2}{\beta}\int{d^3 {\bf
q}\over (2\pi)^3}\ln\left(1-e^{-\beta c|{\bf q}|}\right).
\end{equation}
As we will show below, the former and the latter correspond to the
NBEC and RBEC region, respectively. The bosons and fermions behave
very differently in thermodynamics. As is well known, the specific
heat $C$ of an ideal boson gas is proportional to $T^\alpha$ with
$\alpha=3/2$ and $3$ corresponding to non-relativistic and
ultra-relativistic systems, but the naive BCS mean field theory
predicts an exponential law $C\propto e^{-\Delta_0/T}$, where
$\Delta_0$ is the gap at zero temperature, $\Delta_0=\Delta(T=0)$.

\section {BCS-NBEC-RBEC Crossover with Massive Fermions}
\label{s4}
In this section, we study the BCS-BEC crossover when the coupling
constant $g$ increases. Since our model is non-renormalizable, a
proper regularization is needed. In the case with massive fermions
we employ the often used non-relativistic regularization, namely,
to replace the bare coupling $g$ by a renormalized coupling
$U$\cite{RBCSBEC2,RBCSBEC4},
\begin{equation}
-\frac{1}{U}=\frac{1}{g}-\frac{1}{2}\int_{|{\bf k}|\leq
\Lambda}{d^3{\bf k}\over (2\pi)^3}\left(\frac{1}{\epsilon_{\bf
k}-m}+\frac{1}{\epsilon_{\bf k}+m}\right).
\end{equation}
The effective s-wave scattering length $a_s$ can be related to $U$
by $U=4\pi a_s/m$. While this is a natural extension of the
non-relativistic regularization to relativistic systems, the
ultraviolet divergence can not be completely removed, and a cutoff
$\Lambda$ still exists in the theory. In this regularization, the
solution of the coupled equations depends on three dimensionless
parameters: the effective coupling constant $\eta=1/(k_fa_s)$, the
quantity $\zeta=k_f/m$ which reflects the fermion number density,
and the cutoff $\Lambda/m$.

We assume in this section that the fermion density $n$ is not very
high and satisfies the relation $n<m^3$ or $\zeta<1$. In this case
the system is not ultra-relativistic and can even be treated
non-relativistically in some parameter region. From the study in
NSR theory above $T_c$ and in the BCS-Leggett theory at $T=0$, if
the dimensionless coupling $\eta$ varies from $-\infty$ to
$+\infty$, the system will undergo two
crossovers\cite{RBCSBEC1,RBCSBEC2,RBCSBEC3,RBCSBEC4}, the
crossover from the BCS state to the NBEC state around $\eta\sim 0$
and the crossover from the NBEC state to the RBEC state around
$\eta\sim\zeta^{-1}$. The NBEC state and the RBEC state are
characterized by the molecule binding energy $E_b$. We have
$E_b\ll2m$ in the NBEC state and $E_b\sim 2m$ in the RBEC state.
\\ 1){\bf BCS region.}  In weak coupling BCS region, there exist no bound
pairs in the system. In this case, $Z_1$ is sufficiently small and
$Z_2$ dominates the pair dispersion\cite{BCSBEC7}, and we have
$\Delta_{\text{pg}}^2\propto1/(Z_2c^3)$ after a simple algebra.
Since $\Delta$ should be small in the weak coupling region, and
$c$ can be proven to be approximately equal to the Fermi
velocity\cite{BCSBEC7}, the pseudogap $\Delta_{pg}$ is very small
and can be safely neglected in this region, as we expected.
Therefore, the BCS mean field approximation is good enough at any
temperature, and the critical temperature satisfies the well known
relation $T_c\simeq0.57\Delta_0$. For example, in the
non-relativistic limit with $k_f\ll m$, the anti-fermion degrees
of freedom can be ignored and the pair susceptibility recovers its
non-relativistic version\cite{BCSBEC6}, see the result in Appendix
\ref{app}. The critical temperature can be expressed
as\cite{BCSBEC6}
\begin{equation}
T_c=\frac{8e^{\gamma-2}}{\pi}\epsilon_fe^{2\eta/\pi},
\end{equation}
where $\gamma$ is the Euler constant and $\epsilon_f=k_f^2/(2m)$
is the Fermi kinetic energy. In this region, even though $Z_2$
dominates the pair dispersion, we can show that $c\propto\Delta$
is vanishingly small due to the weak coupling. Since the boson
contribution to thermodynamics can be neglected, the specific heat
at low temperature takes the well known form $C\propto
e^{-\Delta_0/T}$.
\\2){\bf NBEC region.}  In the non-relativistic BEC region with
$\eta>1$ but $\eta\ll \zeta^{-1}$\cite{RBCSBEC4}, the molecule
binding energy $E_b$ is much less than $2m$, namely $|\mu-m|\ll
m$, the boson mass is approximately $2m$, and the system can be
regarded as a non-relativistical boson gas, if $k_f/m$ is small
enough. Assuming $k_f\ll m$, the anti-fermion degrees of freedom
can be neglected, and we can recover the non-relativistic
result\cite{BCSBEC6}. In this region, the gap $\Delta$ becomes as
large as the Fermi kenetic energy $\epsilon_f$. From
$Z_1\propto1/\Delta^2$ and $Z_2\propto1/\Delta^4$, $Z_1$ is the
dominant one and the pair dispersion becomes quadratic in $|{\bf
q}|$. In this case, the propagator of the uncondensed pairs can be
approximated by
\begin{equation}
t_\text{pg}(q)\simeq\frac{-iZ_1^{-1}}{q_0-|{\bf
q}|^2/\left(2m_\text{B}\right)},
\end{equation}
where the pair mass $m_{\text{B}}$ is defined by
$m_{\text{B}}=Z_1/2\xi^2$, and we have the simple relation
\begin{equation}
Z_1\Delta_{\text{pg}}^2=\int\frac{d^3{\bf
q}}{(2\pi)^3}b\left(\frac{|{\bf
q}|^2}{2m_{\text{B}}}\right)=\left(\frac{m_\text{B}
T}{2\pi}\right)^{3/2}\zeta\left(\frac{3}{2}\right).
\end{equation}
Since $Z_1\Delta_{\text{pg}}^2$ is equal to the total boson
density $n_{\text{B}}$ at $T=T_c$, we arrive at the standard
critical temperature for Bose-Einstein condensation in
non-relativistic boson gas\cite{kerson},
\begin{equation}
T_c=\frac{2\pi}{m_{\text B}}\left(\frac{n_{\text
B}}{\zeta(\frac{3}{2})}\right)^{2/3}.
\end{equation}
The boson mass $m_{\text B}$ is generally expected to be equal to
the boson chemical potential $\mu_{\text B}=2\mu$. In the
non-relativistic limit $k_f\ll m$, we can show
$m_{\text{B}}\simeq2m$ and $Z_1\Delta_{\text{pg}}^2\simeq n/2$ at
$T=T_c$, the critical temperature becomes $T_c=0.218\epsilon_f$.
Since $Z_1$ dominates the pair dispersion, the pseudogap is
proportional to $T^{3/4}$ and the specific heat is proportional to
$T^{3/2}$ at low temperature.
\\3){\bf RBEC
region.} In this region we have the molecule binding energy
$E_b\rightarrow 2m$ and chemical potential $\mu\rightarrow 0$. In
this case, non-relativistic limit can not be reached even for
$k_f\ll m$\cite{RBCSBEC4}. Since the bosons with mass
$m_{\text{B}}=2\mu$ become nearly massless in this region, the
anti-bosons and anti-fermions can be easily excited, and the
system contains both bosons and anti-bosons. From the relation
\begin{equation}
n_{\text{B}}=n_{\text b}-n_{\bar{\text b}}=Z_1\Delta_{\text{pg}}^2
\end{equation}
at $T=T_c$, where $n_{\text b}$ and $n_{\bar{\text b}}$ are the
boson and anti-boson numbers, while $n_{\text b}$ and
$n_{\bar{\text b}}$ are both very large, their difference produces
a small pure boson density $n_{\text{B}}\simeq n/2$. On the other
hand, for $\mu\rightarrow 0$ we can expand $Z_1$ in powers of
chemical potential $\mu$,
\begin{equation}
Z_1\simeq R\mu+O(\mu^3)=\frac{R}{2}m_{\text B}+ O(\mu^3)
\end{equation}
and hence $Z_2$ dominates the pair dispersion, which means that
the pseudogap is proportional to $T$ at low temperature. In this
case, the propagator of the uncondensed pairs can be approximated
by
\begin{equation}
t_{\text{pg}}(q)\simeq\frac{-iZ_2^{-1}}{q_0^2-c^2|{\bf q}|^2},
\end{equation}
which leads to the relation
\begin{equation}
Z_2\Delta_{\text{pg}}^2\simeq\int\frac{d^3{\bf
q}}{(2\pi)^3}\frac{b\left(c|{\bf q}|\right)}{c|{\bf
q}|}=\frac{T^2}{12c^3}.
\end{equation}
Combining the above equations, we find
\begin{equation}
T_c=\left(\frac{24c^3Z_2}{R}\frac{n_{\text{B}}}{m_{\text{B}}}\right)^{1/2}.
\end{equation}
In the RBEC limit $\mu\rightarrow0$, we can approach to the
standard critical temperature for ultra-relativistic Bose-Einstein
condensation\cite{kapusta,RBEC},
\begin{equation}
\label{tc}
T_c=\left(\frac{3n_{\text B}}{m_{\text
B}}\right)^{1/2}.
\end{equation}
Since $n_{\text{B}}$ is almost fixed and $m_{\text
B}\rightarrow0$, $T_c$ would approach to infinity in the RBEC
limit. In the ultra-relativistic boson gas, the specific heat at
low temperature is proportional to $T^3$.

\begin{figure}[!htb]
\begin{center}
\includegraphics[width=6cm]{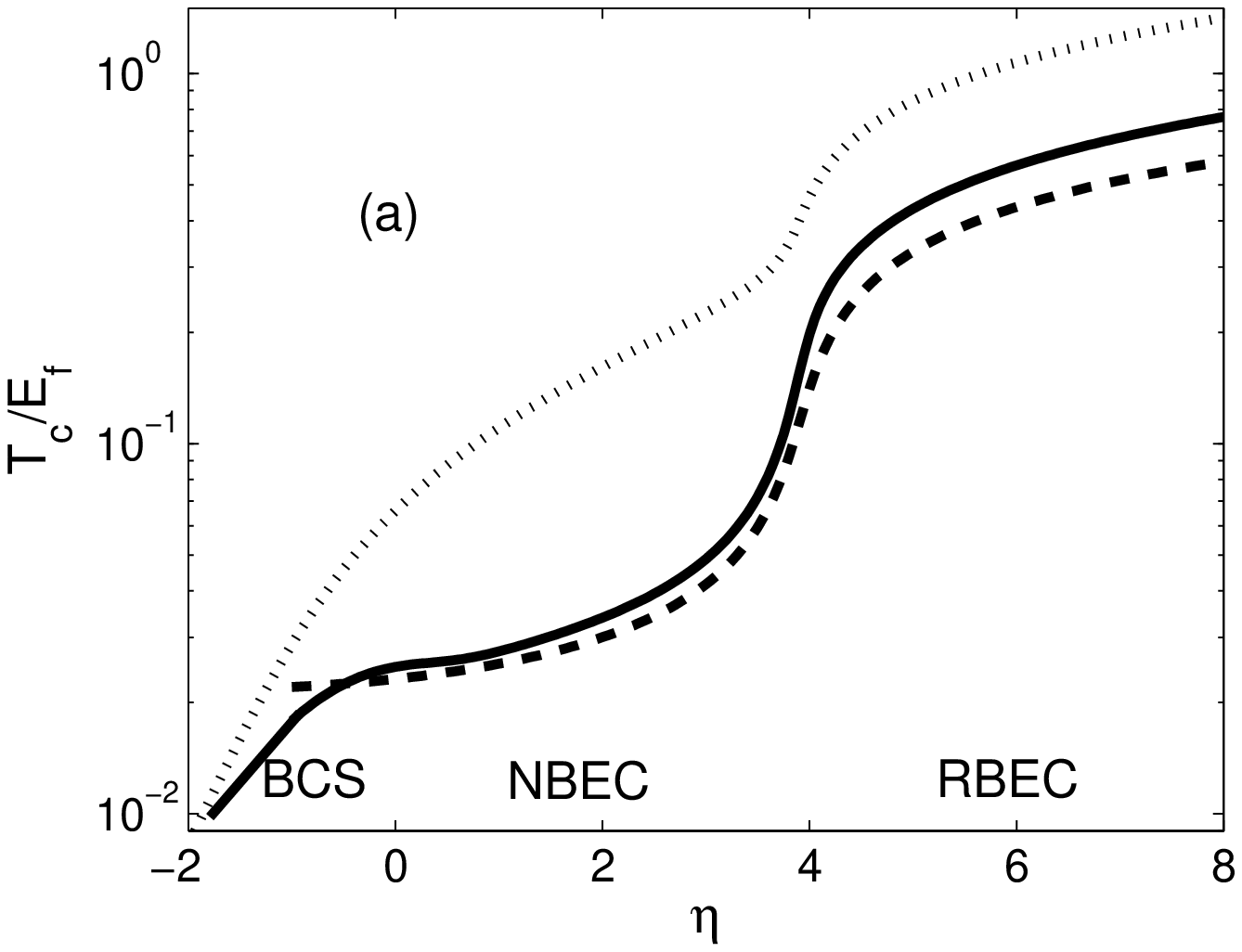}
\includegraphics[width=6cm]{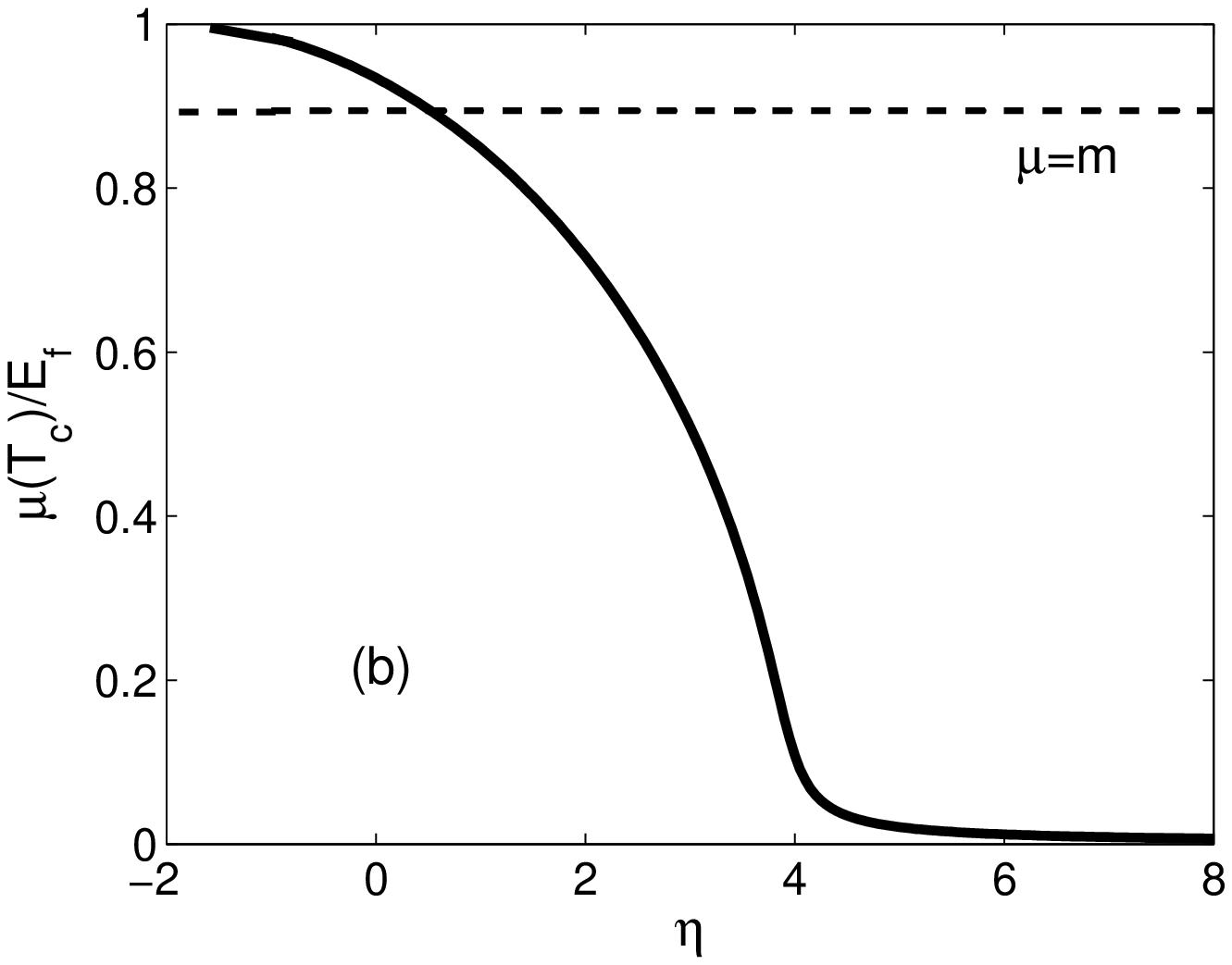}
\includegraphics[width=6cm]{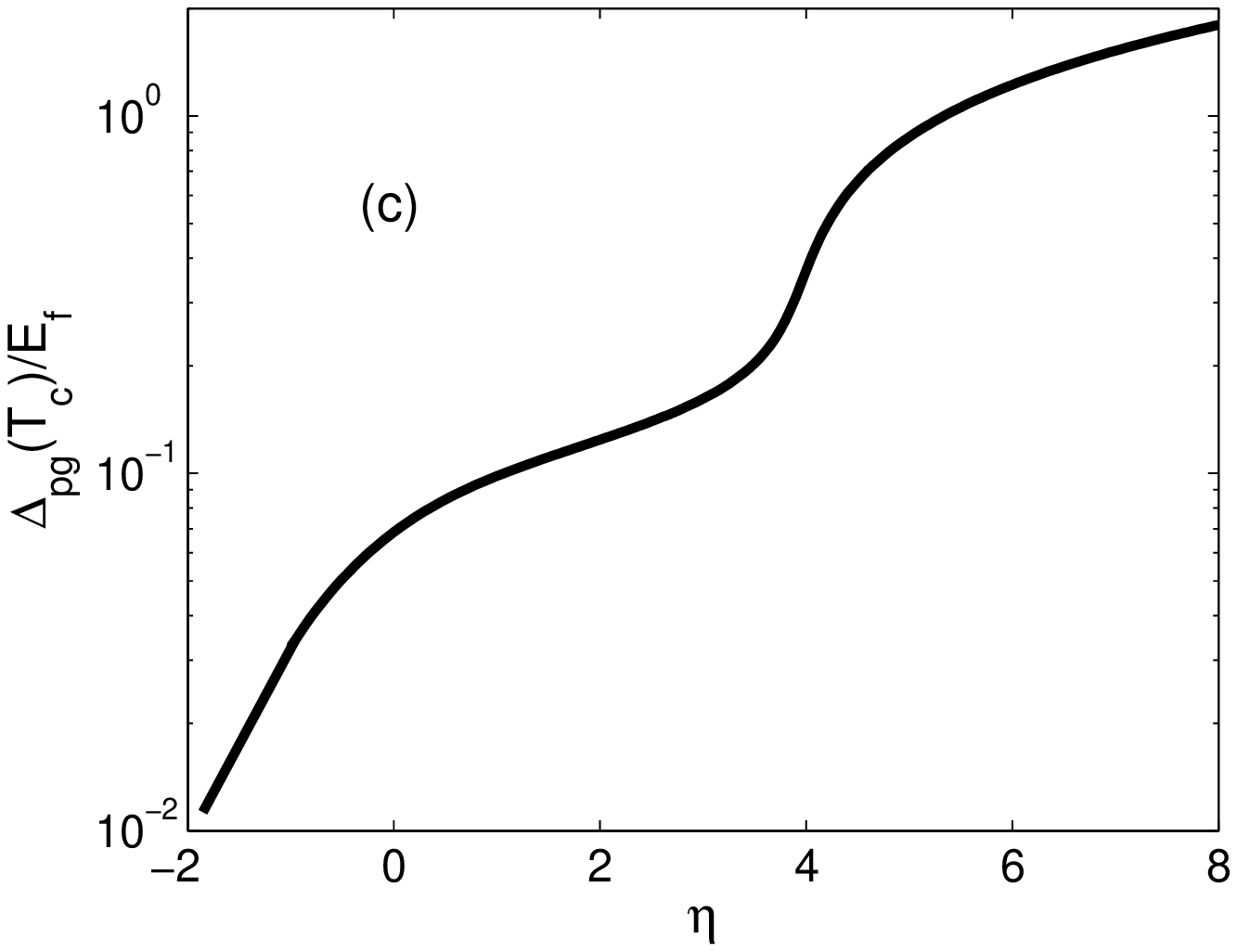}
\caption{The critical temperature $T_c$ (a), chemical potential
$\mu(T_c)$ (b) and pseudogap $\Delta_\text{pg}(T_c)$ (c) as
functions of coupling $\eta$ at $\Lambda/m=10$ and $k_f/m=0.5$.
$T_c, \mu$ and $\Delta_\text{pg}$ are all scaled by the Fermi
energy $E_f$. The dashed line is the standard critical temperature
for the ideal boson gas in (a) and stands for the position $\mu=m$
in (b), and the dotted line in (a) is the limit temperature $T^*$
where the pseudogap starts to disappear. \label{fig2}}
\end{center}
\end{figure}

We now turn to numerical calculations. From the coupled equations
(\ref{gap1}), (\ref{number}) and (\ref{pg2}), we can solve the
critical temperature $T_c$, chemical potential $\mu(T_c)$ and
pseudogap $\Delta_{\text{pg}}(T_c)$ at $T_c$ as functions of the
coupling $\eta$ at fixed $k_f/m$. In Fig.\ref{fig2} we show the
numerical results with the parameters $\Lambda/m=10$ and
$k_f/m=0.5$. The BCS-NBEC-RBEC crossover can be seen directly from
the behavior of the chemical potential $\mu$. In the BCS region
with $-\infty<\eta<0.5$, $\mu$ is larger than the fermion mass and
approaches to the Fermi energy $E_f$ in the weak coupling limit
$\eta\rightarrow-\infty$. The NBEC region is located around
$-0.5<\eta<4$ and the RBEC region is at about $\eta>4$. The
critical coupling $\eta\simeq4$ for the RBEC state is consistent
with our analytical result
\begin{equation}
\eta_c=\frac{2}{\pi}\left(\frac{k_f}{m}\right)^{-1}\ln\left(\Lambda/m+\sqrt{\left(\Lambda/m\right)^2+1}\right)
\end{equation}
derived in \cite{RBCSBEC4}. The difference between NBEC and RBEC
states is that the chemical potential $\mu$ is of the order of $m$
in the NBEC region but approaches zero in the RBEC region.
\begin{figure}[!htb]
\begin{center}
\includegraphics[width=6cm]{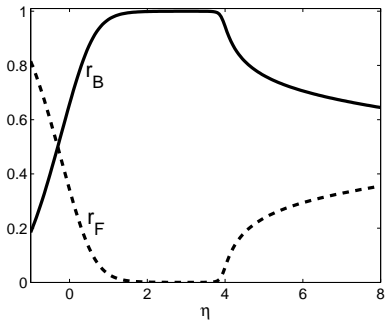}
\caption{The boson number fraction $r_{\text B}$ and the fermion
number fraction $r_{\text F}$ at the critical temperature $T_c$ as
functions of the coupling $\eta$ at $\Lambda/m=10$ and
$k_f/m=0.5$. \label{fig3}}
\end{center}
\end{figure}

The critical temperature, plotted as the solid line in
Fig.\ref{fig2}a, shows significant change from the weak to strong
coupling. To compare it with the standard critical temperature for
the idea boson gas, we solve the equation\cite{kapusta}
\begin{equation}
\int\frac{d^3{\bf q}}{(2\pi)^3}\left[b\left(\epsilon_{\bf
q}^{\text B}-\mu_{\text B}\right)-b\left(\epsilon_{\bf q}^{\text
B}+\mu_{\text B}\right)\right]\Big|_{\mu_{\text B}=m_{\text
B}}=n_{\text B}
\end{equation}
with $\epsilon_{\bf q}^{\text B}=\sqrt{{\bf q}^2+m_{\text B}^2}$,
boson mass $m_{\text B}=2\mu$ and boson number $n_{\text B}=n/2$,
and show the obtained critical temperature as dashed line in
Fig.\ref{fig2}a. In the weak coupling region $T_c$ is very small
and agrees with the BCS theory. In the NBEC region $T_c$ changes
smoothly and there is no remarkable difference between the solid
and dashed lines. Around the coupling $\eta_c=4$, $T_c$ increases
rapidly and then varies smoothly again. In the RBEC region, the
critical temperature from our calculation deviates significantly
from the standard critical temperature for ideal boson gas. Note
that, $T_c$ is of the order of the Fermi kinetic energy
$\epsilon_f\simeq k_f^2/(2m)$ in the NBEC region but becomes as
large as the Fermi energy $E_f$ in the RBEC region. The pseudogap
$\Delta_\text{pg}$ at $T=T_c$, shown in Fig.\ref{fig2}c, behaves
similarly as the critical temperature. To see clearly the
pseudogap region, we present in Fig.\ref{fig2}a the limit
temperature $T^*$ as a dotted line. The pseudogap exists between
the solid and dotted lines and begin to vanish on the dotted line.

To explain why the critical temperature in the RBEC region
deviates remarkably from the standard one for ideal boson gas, we
calculate the boson number fraction $r_{\text B}=n_{\text
B}/(n/2)$ and the fermion number fraction $r_{\text F}=1-r_{\text
B}$ and show them as functions of the coupling $\eta$ in
Fig.\ref{fig3}. While there are only bosons at $T_c$ in the NBEC
region, $r_{\text B}$ is obviously less than $1$ in the RBEC
region. This conclusion is consistent with the results from the
NSR theory\cite{RBCSBEC1,RBCSBEC2}. In the NBEC region, the
binding energy of the molecules is $E_b\simeq
1/ma_s^2=2\eta^2\epsilon_f$, which is much larger than the
critical temperature $T_c\simeq0.2\epsilon_f$, and the molecules
can be safely regarded as point bosons at temperature near $T_c$.
However, the critical temperature in the RBEC region is as large
as the Fermi energy $E_f$, which is of the order of the molecule
binding energy $E_b\simeq2m$. Due to the competition between the
condensation and dissociation of composite bosons in hot medium,
the molecules can not be regarded as point bosons and the critical
temperature should deviates from the result for ideal boson gas.
This may be a general characteristic of a composite boson system,
when the condensation temperature $T_c$ is of the order of the
molecule binding energy. The phenomenon can be explained by the
competition between free energy and entropy\cite{RBCSBEC2}: In
terms of entropy a two-fermion state is more favorable than a
one-boson state, but in terms of free energy it is less favorable.
Since the condensation temperature $T_c$ in the RBEC region is of
the order of $(n_{\text B}/m_{\text B})^{1/2}\sim(n/\mu)^{1/2}$,
we conclude that only for a system with sufficiently small value
of $k_f/m$, the standard RBEC critical temperature can be reached
and is much smaller than $2m$.

\section {Application To Massless Fermions: Color Superconductivity}
\label{s5}
As a natural application of the relativistic $G_0 G$ formalism, we
calculate in this section the transition temperature and pseudogap
in two flavor color superconductivity at moderate baryon density.
The two flavor color superconducting quark matter corresponds to
the ultra-relativistic case with $n\gg m^3$, where $m$ is the
current quark mass. At moderate baryon density, the quark energy
gap due to color superconductivity is of the order of 100 MeV,
which is not located in the weak coupling region. As a result, the
pseudogap effect is expected to be significantly important near
the critical temperature. To apply the present theory directly, we
employ the generalized Nambu--Jona-Lasinio(NJL) model with scalar
diquark channel, which has been widely used to study color
superconductivity at moderate baryon density. The Lagrangian
density is defined as
\begin{eqnarray}
{\cal L} &=&
\bar{\psi}\left(i\gamma^{\mu}\partial_{\mu}-m\right)\psi
+G_{\text{s}}\left[\left(\bar{\psi}\psi\right)^2+\left(\bar{\psi}i\gamma_5
\tau\psi\right)^2 \right]\\
&&+G_{\text{d}}\sum_{a=2,5,7}\left(\bar\psi
i\gamma^5\tau_2\lambda_a C\bar{\psi}^{\text
T}\right)\left(\psi^{\text T}C
i\gamma^5\tau_2\lambda_a\psi\right),\nonumber
\end{eqnarray}
where $\psi$ and $\bar{\psi}$ denote the quark fields with two
flavors ($N_f=2$) and three colors ($N_c=3$), $\tau_i (i=1,2,3)$
are the Pauli matrices in flavor space and $\lambda_a
(a=1,2,...,8)$ are the Gell-Mann matrices in color space, and
$G_{\text s}$ and $G_{\text d}$ are coupling constants in meson
and diquark channels.

At moderate baryon density, the chiral symmetry has already been
restored and we need not consider the chiral condensate
$\langle\bar{\psi}\psi\rangle$. Since the current quark mass $m$
is about $5$ MeV, the quarks are nearly massless. The order
parameter field for color superconductivity is defined as
\begin{equation}
\Phi_a=-2G_{\text d}\psi^{\text T}C i\gamma^5\tau_2\lambda_a\psi.
\end{equation}
To simplify the calculation, one usually considers a spontaneous
color breaking from the SU(3) symmetry to a SU(2) subgroup. Due to
the residual color SU(2) symmetry, the effective potential in mean
field approximation should depend only on the combination
$\Delta_2^2+\Delta_5^2+\Delta_7^2$ with
$\Delta_a=\langle\Phi_a\rangle$, and we can choose a specific
gauge $\Delta_\text{sc}=\Delta_2\neq 0, \Delta_5=\Delta_7=0$
without loss of generality. In this gauge, the red and green
quarks participate in the condensation, but the blue one does not.

The detailed formalism of the $G_0G$ theory in the NJL model is
similar to what we shown in sections \ref{s2} and \ref{s3} but
becomes somewhat complicated due to the presence of color and
flavor degrees of freedom. Comparing the quark propagator in the
NJL model with the one shown in above sections, the relativistic
$G_0 G$ scheme can be directly applied to the study of color
superconductivity, provided that we consider carefully the
difference between the pairing including a blue quark and the
pairing with only red and green quarks. The dispersion for red and
green quarks is identical with the one obtained in the toy model,
their excitation gap is
$\Delta=(\Delta_\text{sc}^2+\Delta_\text{pg}^2)^{1/2}$, and the
pair susceptibility $\chi(q)$ should be multiplied by a factor
$N_f(N_c-1)$ where $N_f$ and $N_c$ are flavor and color numbers of
quarks. The new feature is that a gapless blue quark in the naive
BCS mean field theory obtains a gap $\Delta_{\text{pg}}$ in the
$G_0 G$ scheme. This can be understood by the fact that, the color
symmetry is controlled only by the order parameters themselves,
and fluctuations of any order parameter field $\Phi_a$ do not
change it. At and above the critical temperature,
$\Delta_{\text{sc}}=0$, the color symmetry is restored, all colors
become degenerate, and their gaps are just the pseudogap.

The two flavor quark matter may exist in the region of
$\mu=350-500$ MeV, where the strange quarks are not yet excited.
Unlike the study in above sections in the canonical ensemble with
fixed fermion number, people usually investigate color
superconductivity in the grand canonical ensemble with fixed quark
chemical potential. In this case, the quark number is not directly
coupled to the calculation of the order parameter
$\Delta_\text{sc}$ and pseudogap $\Delta_\text{pg}$. For numerical
calculations, we take the current quark mass $m=5$ MeV, the often
used quark momentum cutoff $\Lambda=650$ MeV, and a fixed quark
chemical potential $\mu=400$ MeV. We have checked that a
reasonable change in the value of $\mu$ does not bring qualitative
difference. As is conventionally considered in the literatures, we
use the pairing gap $\Delta_0$ at zero temperature to reflect the
strength of the diquark coupling constant $G_{\text{d}}$.

\begin{figure}[!htb]
\begin{center}
\includegraphics[width=6cm]{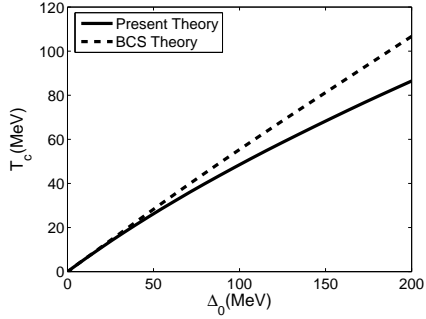}
\caption{The phase transition temperature $T_c$ for two flavor
color superconductivity as a function of the diquark condensate
$\Delta_0$ at zero temperature in the BCS mean field theory
(dashed line) and in the $G_0 G$ theory with diquark fluctuations
(solid line). \label{fig4}}
\end{center}
\end{figure}

\begin{figure}[!htb]
\begin{center}
\includegraphics[width=6cm]{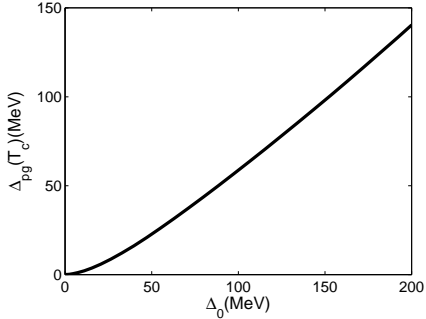}
\caption{The pseudogap $\Delta_{\text{pg}}$ in two flavor color
superconductivity at the critical temperature $T_c$ as a function
of $\Delta_0$. \label{fig5}}
\end{center}
\end{figure}

\begin{figure}[!htb]
\begin{center}
\includegraphics[width=6cm]{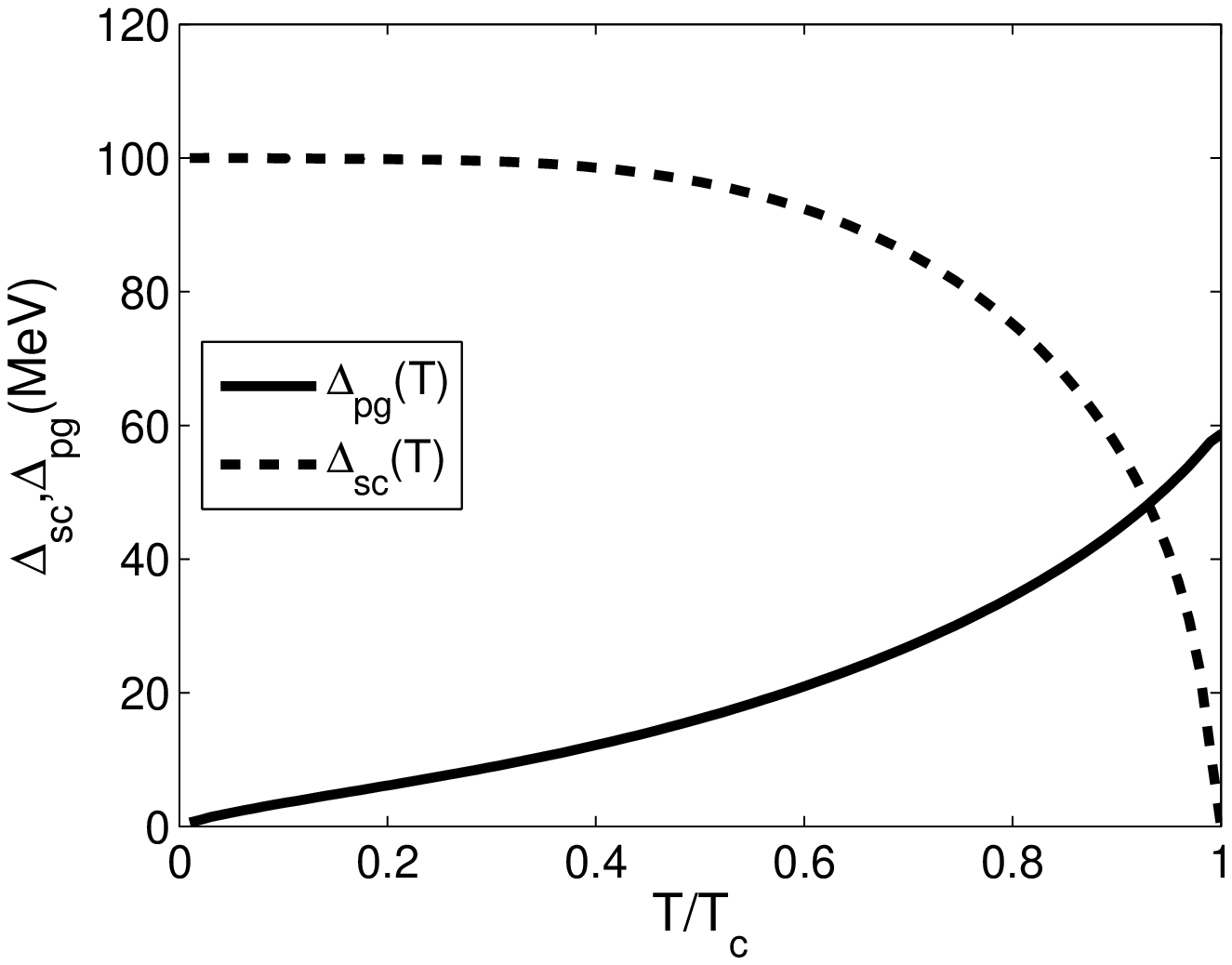}
\includegraphics[width=6cm]{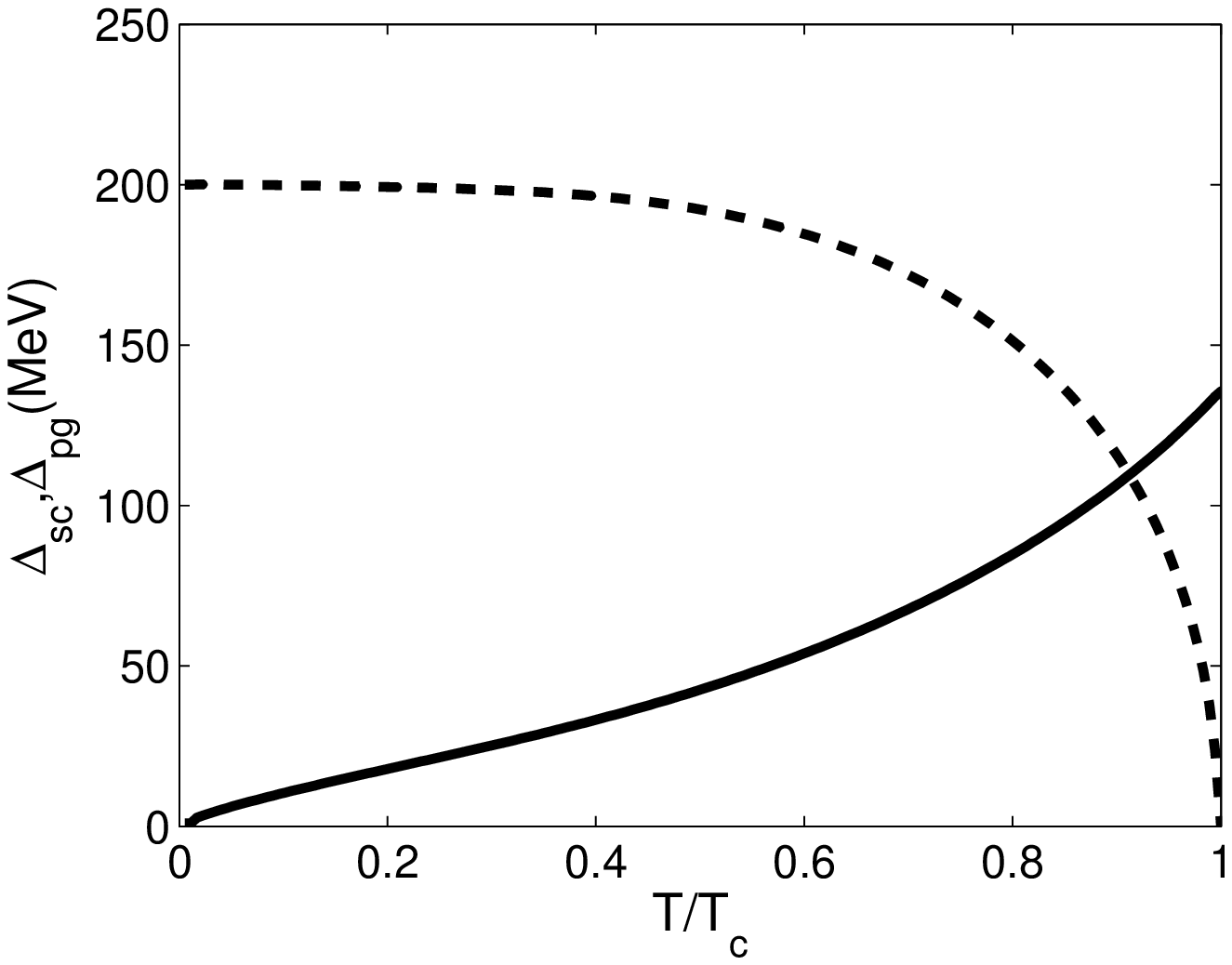}
\caption{The diquark condensate $\Delta_\text{sc}$ (dashed lines)
and pseudogap $\Delta_\text{pg}$ (solid lines) in two flavor color
superconductivity as functions of temperature scaled by $T_c$ for
$\Delta_0=100$ MeV (upper panel) and $200$ MeV (lower panel).
\label{fig6}}
\end{center}
\end{figure}

In Fig.\ref{fig4} we show the critical temperature $T_c$ as a
function of $\Delta_0$ in the $G_0 G$ theory and in the BCS mean
field theory. While the critical temperature is not strongly
modified by the diquark fluctuations in a wide range of
$\Delta_0$, the difference between the two is up to $20\%$ in the
strong coupling region with $\Delta_0\simeq 200$ MeV. In
Fig.\ref{fig5}, we show the pseudogap $\Delta_{\text{pg}}$ at the
critical temperature $T_c$. In a wide range of the coupling, the
pseudogap is of the order of $100$ MeV, which is as large as the
diquark condensate $\Delta_\text{sc}$ at zero temperature. Such a
behavior means that the two flavor color superconductivity at
moderate density is in the BCS-BEC crossover region and quite like
the high temperature superconductivity in
cuprates\cite{BCSBEC6,BCSBEC7}. Since $\Delta_\text{sc}$ vanishes
at $T=T_c$, the large pseudogap will bring significant effect at
and above $T_c$, such as the non-Fermi liquid behavior. In
Fig.\ref{fig6}, we show the temperature dependence of the diquark
condensate $\Delta_\text{sc}$ and pesudogap $\Delta_\text{pg}$ at
two values of $\Delta_0$. With increasing temperature, while the
diquark condensate decreases, the pseudogap increases from zero.
At low temperature, especially at zero temperature, we can safely
neglect the pseudogap.
\begin{figure}[!htb]
\begin{center}
\includegraphics[width=6cm]{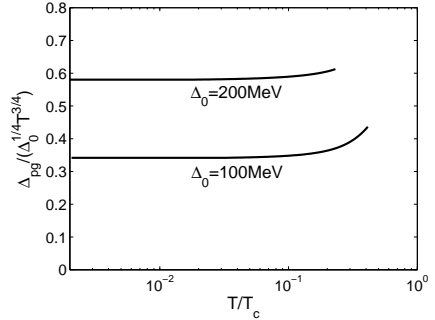}
\caption{The temperature dependence of the pseudogap
$\Delta_\text{pg}$ scaled by $\Delta_0^{1/4}T^{3/4}$ for
$\Delta_0=100$ MeV and $200$ MeV. \label{fig7}}
\end{center}
\end{figure}

While the pseudogap is small at low temperature and dominates the
system only near and above $T_c$, the diquark fluctuations bring
significant contribution to thermodynamics at any temperature. In
the low temperature region, the temperature behavior of the
pseudogap is significantly important, since it can tell us whether
the coefficient $Z_1$ or $Z_2$ dominates the pair fluctuations. In
Fig.\ref{fig7} we show the pseudogap at low temperature. In the
region of $T/T_c\leq 0.1$, it obeys a perfect power law
$\Delta_\text{pg}\propto T^{3/4}$, which means that $Z_1$ is the
dominant one for the pair susceptibility.

Considering the uncondensed diquarks, the total thermodynamic
potential $\Omega$ can be expressed as
\begin{equation}
\Omega=\Omega_\text{cond}+\Omega_\text{quark}+\Omega_\text{diquark},
\end{equation}
where the condensate and quark contributions take the same form as
in the BCS theory, and the diquark contribution can be written as
\begin{equation}
\Omega_{\text{diquark}}=\sum_q\ln[1-4G_\text{d}\chi(q)].
\end{equation}
Since the coefficient $Z_1$ controls the pair fluctuations at low
temperature, the specific heat satisfies the power law $C\propto
T^{3/2}$. As we mentioned above, the diquark contribution can be
neglected only at sufficiently weak coupling. While the color
superconductor at moderate baryon density may not reach the BEC
condition, the effect of diquark fluctuations on the
thermodynamics may be remarkable, and it may bring significant
astrophysical consequences, such as the cooling process in compact
stars.

\section {Summary}
\label{s6}
We have generalized the non-relativistic $G_0 G$ formalism of
BCS-BEC crossover to relativistic fermion systems. The theory can
describe the superfluidity/superconductivity with arbitrary
strength of attractive interaction, both in the symmetric phase
and symmetry breaking phase. The beyond-BCS effect at strong
coupling brings in thermally excited bosons and contributes a
pseudogap to fermion excitations. In such a formalism, we
confirmed that there exists a BCS-NBEC-RBEC crossover in
relativistic fermion systems.

For color superconductivity at moderate baryon density, while the
BEC state can not be reached, the effect of diquark fluctuations
is still remarkable and the naive BCS mean field theory breaks
down when the temperature is close to the critical value. We
investigate the two flavor color superconductivity at quark
chemical potential $\mu=350-500$ MeV where the gap at zero
temperature is of the order of 100 MeV. We found that the
beyond-BCS effect strongly suppresses the transition temperature,
and the pseudogap is very large near the critical temperature.
This may strongly modify the thermodynamics of quark matter and
bring significant astrophysical consequences in the study of
compact stars.

Such a theory can be applied to not only diquark condensate
($\langle qq\rangle$) at finite baryon density but also chiral
condensate ($\langle q\bar q\rangle$) at finite temperature and
pion superfluidity at finite isospin density. The observation of
$q\bar q$ bound states in strongly coupled quark-gluon
plasma\cite{SQGP} and a large thermal quark mass above the chiral
phase transition temperature in lattice QCD\cite{LAT} indicate
strongly the significance of the $q\bar{q}$ Bose-Einstein
condensation and the quark pseudogap effect\cite{qqbar,qqbar2}.
The study in this direction is under progress\cite{new}.

{\bf Acknowledgement:} We thank Dr.Mei Huang for drawing our
attention to \cite{BCSBEC6} and \cite{BCSBEC7} and Gaofeng Sun for
a careful check on the calculations in the Appendix. This work is
supported by the grants No.NSFC10428510, No.10575058 and the major
state basic research developing program 2007CB815000.

\appendix
\begin{widetext}
\section{ Pair Susceptibility and Its Expansion Coefficients}
\label{app}
In this appendix, we evaluate the explicit expression of the pair
susceptibility and its momentum expansion. Completing the trace in
Dirac space and the Matsubara summation over the fermion
frequencies, we obtain from the equations (\ref{sus}),
(\ref{element}) and (\ref{g0})
\begin{eqnarray}
\chi(q)&=&\int\frac{d^3{\bf
k}}{(2\pi)^3}\Bigg[\left(\frac{1-f(E_{\bf k}^-)-f(\xi_{{\bf
q}-{\bf k}}^-)}{E_{\bf k}^-+\xi_{{\bf q}-{\bf k}}^--q_0}
\frac{E_{\bf k}^-+\xi_{\bf k}^-}{2E_{\bf k}^-}-\frac{f(E_{\bf
k}^-)-f(\xi_{{\bf q}-{\bf k}}^-)} {E_{\bf k}^--\xi_{{\bf q}-{\bf
k}}^-+q_0}\frac{E_{\bf k}^--\xi_{\bf k}^-}{2E_{\bf k}^-}\right)
\left(\frac{1}{2}-\frac{{\bf k}\cdot({\bf q}-{\bf
k})-m^2}{2\epsilon_{\bf k}\epsilon_{{\bf q}-{\bf k}}}\right)
\nonumber\\
&&+\left(\frac{1-f(E_{\bf k}^-)-f(\xi_{{\bf q}-{\bf k}}^+)}{E_{\bf
k}^-+\xi_{{\bf q}-{\bf k}}^++q_0} \frac{E_{\bf k}^--\xi_{\bf
k}^-}{2E_{\bf k}^-}-\frac{f(E_{\bf k}^-)-f(\xi_{{\bf q}-{\bf
k}}^+)} {E_{\bf k}^--\xi_{{\bf q}-{\bf k}}^+-q_0}\frac{E_{\bf
k}^-+\xi_{\bf k}^-}{2E_{\bf k}^-}\right)
\left(\frac{1}{2}+\frac{{\bf k}\cdot({\bf q}-{\bf
k})-m^2}{2\epsilon_{\bf k}\epsilon_{{\bf q}-{\bf k}}}\right)\Bigg]
\nonumber\\
&&+\left(E_{\bf k}^\pm,\xi_{\bf k}^\pm,q_0\rightarrow E_{\bf
k}^\mp,\xi_{\bf k}^\mp,-q_0\right).
\end{eqnarray}
Taking its first and second order derivatives with respect to
$q_0$, we have
\begin{eqnarray}
Z_1&=&\int\frac{d^3{\bf k}}{(2\pi)^3}\frac{1}{2E_{\bf
k}^-}\left[\frac{1-f(E_{\bf k}^-)-f(\xi_{{\bf k}}^-)}{E_{\bf
k}^-+\xi_{\bf k}^-}+\frac{f(E_{\bf k}^-)-f(\xi_{\bf k}^-)}{E_{\bf
k}^--\xi_{\bf k}^-}\right]-\left(E_{\bf k}^\pm,\xi_{\bf k}^\pm\rightarrow E_{\bf k}^\mp,\xi_{\bf k}^\mp\right),\nonumber\\
Z_2&=&\int\frac{d^3{\bf k}}{(2\pi)^3}\frac{1}{2E_{\bf
k}^-}\left[\frac{1-f(E_{\bf k}^-)-f(\xi_{{\bf k}}^-)}{(E_{\bf
k}^-+\xi_{\bf k}^-)^2}-\frac{f(E_{\bf k}^-)-f(\xi_{\bf
k}^-)}{(E_{\bf k}^--\xi_{\bf k}^-)^2}\right]+\left(E_{\bf
k}^\pm,\xi_{\bf k}^\pm\rightarrow E_{\bf k}^\mp,\xi_{\bf
k}^\mp\right).
\end{eqnarray}
Using the relation $(E_{\bf k}^\pm)^2-(\xi_{\bf
k}^\pm)^2=\Delta^2$, the coefficients can be rewritten as
\begin{eqnarray}
Z_1&=&\frac{1}{2\Delta^2}\left[n-2\int\frac{d^3{\bf
k}}{(2\pi)^3}\left(f(\xi_{\bf k}^-)-f(\xi_{\bf
k}^+)\right)\right],\nonumber\\
Z_2&=&\frac{1}{2\Delta^4}\int\frac{d^3{\bf
k}}{(2\pi)^3}\left[\frac{(E_{\bf k}^-)^2+(\xi_{\bf k}^-)^2}{E_{\bf
k}^-}\left(1-2f(E_{\bf k}^-)\right)-2\xi_{\bf
k}^-\left(1-2f(\xi_{\bf k}^-)\right)\right]+\left(E_{\bf
k}^\pm,\xi_{\bf k}^\pm\rightarrow E_{\bf k}^\mp,\xi_{\bf
k}^\mp\right).
\end{eqnarray}
Taking the second order derivative of the susceptibility $\chi$
with respect to ${\bf q}$, we obtain another coefficient
\begin{eqnarray}
\xi^2&=&\frac{1}{2}\int\frac{d^3{\bf
k}}{(2\pi)^3}\Bigg[\frac{1}{2E_{\bf k}^-}\left(\frac{1-f(E_{\bf
k}^-)-f(\xi_{{\bf k}}^-)}{E_{\bf k}^-+\xi_{\bf
k}^-}+\frac{f(E_{\bf k}^-)-f(\xi_{\bf k}^-)}{E_{\bf k}^--\xi_{\bf
k}^-}\right)\frac{\epsilon_{\bf k}^2-{\bf k}^2x^2}{\epsilon_{\bf
k}^3}\nonumber\\
&&-\left(\frac{1}{E_{\bf k}^-}\left(\frac{1-f(E_{\bf
k}^-)-f(\xi_{{\bf k}}^-)}{(E_{\bf k}^-+\xi_{\bf
k}^-)^2}-\frac{f(E_{\bf k}^-)-f(\xi_{\bf k}^-)}{(E_{\bf
k}^--\xi_{\bf k}^-)^2}\right)+\frac{2f^\prime(\xi_{\bf
k}^-)}{\Delta^2}\right)\frac{{\bf k}^2x^2}{\epsilon_{\bf k}^2}\nonumber\\
&&-\left(\frac{1-f(E_{\bf k}^-)-f(\xi_{\bf k}^+)}{E_{\bf
k}^-+\xi_{\bf k}^+}\frac{E_{\bf k}^--\xi_{\bf k}^-}{2E_{\bf
k}^-}-\frac{f(E_{\bf k}^-)-f(\xi_{\bf k}^+)}{E_{\bf k}^--\xi_{\bf
k}^+}\frac{E_{\bf k}^-+\xi_{\bf k}^-}{2E_{\bf
k}^-}-\frac{1-2f(E_{\bf k}^-)}{2E_{\bf
k}^-}\right)\frac{\epsilon_{\bf k}^2-{\bf k}^2x^2}{2\epsilon_{\bf
k}^4}\Bigg]\nonumber\\
&&+\left(E_{\bf k}^\pm,\xi_{\bf k}^\pm\rightarrow E_{\bf
k}^\mp,\xi_{\bf k}^\mp\right)
\end{eqnarray}
with $x=\cos\theta$ and $f^\prime(x)$ being the first order
derivative of the Fermi-Dirac distribution.

In the non-relativistic limit with $k_f\ll m, |\mu-m|, \Delta\ll
m$, all the terms including anti-fermion dispersions can be safely
neglected, and the relativistic dispersions are reduced to
$\xi_{\bf k}={\bf k}^2/(2m)-(\mu-m)$ and $E_{\bf k}=\sqrt{\xi_{\bf
k}^2+\Delta^2}$. Taking into account $|{\bf q}|\ll m$, we have
\begin{eqnarray}
\chi_{\text{NR}}(q)&=&\int\frac{d^3{\bf
k}}{(2\pi)^3}\left[\frac{1-f(E_{\bf k})-f(\xi_{{\bf q}-{\bf
k}})}{E_{\bf k}+\xi_{{\bf q}-{\bf k}}-q_0}\frac{E_{\bf k}+\xi_{\bf
k}}{2E_{\bf k}}-\frac{f(E_{\bf k})-f(\xi_{{\bf q}-{\bf
k}})}{E_{\bf k}-\xi_{{\bf q}-{\bf k}}+q_0}\frac{E_{\bf k}-\xi_{\bf
k}}{2E_{\bf k}}\right],
\end{eqnarray}
which is just the same as the one given in \cite{BCSBEC6,BCSBEC7},
and the expansion coefficients $Z_1, Z_2$ and $\xi^2$ are reduced
to\cite{BCSBEC6,BCSBEC7}
\begin{eqnarray}
Z_1&=&\int\frac{d^3{\bf k}}{(2\pi)^3}\frac{1}{2E_{\bf
k}}\left[\frac{1-f(E_{\bf k})-f(\xi_{{\bf k}})}{E_{\bf k}+\xi_{\bf
k}}+\frac{f(E_{\bf k})-f(\xi_{\bf k})}{E_{\bf
k}-\xi_{\bf k}}\right]\nonumber\\
&=&\frac{1}{2\Delta^2}\left[n-2\int\frac{d^3{\bf
k}}{(2\pi)^3}f(\xi_{\bf k})\right],\nonumber\\
Z_2&=&\int\frac{d^3{\bf k}}{(2\pi)^3}\frac{1}{2E_{\bf
k}}\left[\frac{1-f(E_{\bf k})-f(\xi_{{\bf k}})}{(E_{\bf
k}+\xi_{\bf k})^2}-\frac{f(E_{\bf k})-f(\xi_{\bf k})}{(E_{\bf
k}-\xi_{\bf k})^2}\right]\nonumber\\
&=&\frac{1}{2\Delta^4}\int\frac{d^3{\bf
k}}{(2\pi)^3}\left[\frac{E_{\bf k}^2+\xi_{\bf k}^2}{E_{\bf
k}}\left(1-2f(E_{\bf k})\right)-2\xi_{\bf k}\left(1-2f(\xi_{\bf
k})\right)\right],\nonumber\\
\xi^2&=&\int\frac{d^3{\bf k}}{(2\pi)^3}\Bigg[\frac{1}{4mE_{\bf
k}}\left(\frac{1-f(E_{\bf k})-f(\xi_{{\bf k}})}{E_{\bf k}+\xi_{\bf
k}}+\frac{f(E_{\bf k})-f(\xi_{\bf k})}{E_{\bf
k}-\xi_{\bf k}}\right)\nonumber\\
&&-\frac{{\bf k}^2}{6m^2}\left(\frac{1}{E_{\bf
k}}\left(\frac{1-f(E_{\bf k})-f(\xi_{{\bf k}})}{(E_{\bf
k}+\xi_{\bf k})^2}-\frac{f(E_{\bf k})-f(\xi_{\bf k})}{(E_{\bf
k}-\xi_{\bf k})^2}\right)+\frac{2f^\prime(\xi_{\bf
k})}{\Delta^2}\right)\Bigg].
\end{eqnarray}

In the RBEC limit with $\mu\rightarrow 0$, we can expand $Z_1$ in
powers of $\mu$, $Z_1\simeq R\mu+O(\mu^3)$, with the expansion
coefficient $R$ given by
\begin{equation}
R=\int\frac{d^3{\bf k}}{(2\pi)^3}\left[\frac{1-2f(E_{\bf
k})}{E_{\bf k}^3}-2\frac{\epsilon_{\bf k}^2}{E_{\bf
k}^2}\frac{f^\prime(E_{\bf
k})}{\Delta^2}+2\frac{f^\prime(\epsilon_{\bf
k})}{\Delta^2}\right],
\end{equation}
where $E_{\bf k}=\sqrt{\epsilon_{\bf k}^2+\Delta^2}$ is the
dispersion at $\mu=0$, and $Z_2$ and $\xi^2$ can be simplified as
\begin{eqnarray}
Z_2&=&\frac{1}{\Delta^4}\int\frac{d^3{\bf
k}}{(2\pi)^3}\left[\frac{E_{\bf k}^2+\epsilon_{\bf k}^2}{E_{\bf
k}}(1-2f(E_{\bf k}))-2\epsilon_{\bf k}(1-2f(\epsilon_{\bf
k}))\right],\nonumber\\
\xi^2&=&\int\frac{d^3{\bf
k}}{(2\pi)^3}\Bigg[\frac{1}{2\Delta^2}\left(\left(1-2f(\epsilon_{\bf
k})\right)-\frac{\epsilon_{\bf k}}{E_{\bf k}}\left(1-2f(E_{\bf
k})\right)\right)\frac{\epsilon_{\bf k}^2-{\bf
k}^2x^2}{\epsilon_{\bf
k}^3}\nonumber\\
&&-\left(\frac{1}{\Delta^4}\left(\frac{E_{\bf k}^2+\epsilon_{\bf
k}^2}{E_{\bf k}}\left(1-2f(E_{\bf k})\right)-2\epsilon_{\bf
k}\left(1-2f(\epsilon_{\bf
k})\right)\right)+\frac{2f^\prime(\epsilon_{\bf
k})}{\Delta^2}\right)\frac{{\bf k}^2x^2}{\epsilon_{\bf k}^2}\nonumber\\
&&-\left(\frac{1}{2\Delta^2}\left(\frac{E_{\bf k}^2+\epsilon_{\bf
k}^2}{E_{\bf k}}\left(1-2f(E_{\bf k})\right)-2\epsilon_{\bf
k}\left(1-2f(\epsilon_{\bf k})\right)\right)-\frac{1-2f(E_{\bf
k})}{2E_{\bf k}}\right)\frac{\epsilon_{\bf k}^2-{\bf
k}^2x^2}{2\epsilon_{\bf k}^4}\Bigg].
\end{eqnarray}
\end{widetext}

\end{document}